\documentclass[prb, twocolumn, superscriptaddress, aps]{revtex4}
\usepackage{amsmath,amsfonts, amssymb, amsthm, dsfont}
\usepackage[mathscr]{euscript}
\usepackage{bm}
\usepackage{mathrsfs}
\usepackage{graphicx}
\usepackage{verbatim}
\usepackage[colorlinks=true,citecolor=blue]{hyperref}

\newcommand{\di}{\mathrm{d}}

\renewcommand{\vec}[1]{{\mathbf #1}}

\newcommand{\comments}[1]{}
\newcommand{\Ref}[1]{Ref. [\onlinecite{#1}]}

\newcommand{\eq}[1]{Eq.~(\ref{#1})}
\newcommand{\mb}[1]{\mathbf{#1}}

\usepackage[usenames,dvipsnames]{color}         
\definecolor{Zcolour}{rgb}{0.992, 0.588, 0.22}
\definecolor{dkgreen}{rgb}{0,0.5,0}
\definecolor{purple}{rgb}{0.5,0,0.5}

\begin{document}

\title{Topological Quantum Field Theory for Abelian Topological Phases and Loop Braiding Statistics in $(3+1)$-Dimensions}
 
\author{Qing-Rui Wang}
\affiliation{Department of Physics, The Chinese University of Hong Kong, Shatin, New Territories, Hong Kong}

\author{Meng Cheng}
\affiliation{Department of Physics, Yale University, New Haven, CT 06511-8499, USA}

\author{Chenjie Wang}
\affiliation{Department of Physics, City University of Hong Kong, Kowloon, Hong Kong}

\author{Zheng-Cheng Gu}
\affiliation{Department of Physics, The Chinese University of Hong Kong, Shatin, New Territories, Hong Kong}
 

\begin{abstract}
Topological qauntum field theory(TQFT) is a very powerful theoretical tool to study topological phases and phase transitions. In $2+1$D, it is well known that the Chern-Simons theory captures all the universal topological data of topological phases, e.g., quasi-particle braiding statistics, chiral central charge and even provides us a deep insight for the nature of topological phase transitions. Recently, topological phases of quantum matter are also intensively studied in $3+1$D and it has been shown that loop like excitation obeys the so-called three-loop-braiding statistics. In this paper, we will try to establish a TQFT framework to understand the quantum statistics of particle and loop like excitation in $3+1$D. We will focus on Abelian topological phases for simplicity, however, the general framework developed here is not limited to Abelian topological phases. 
\end{abstract}

\maketitle

\tableofcontents

\section{Introduction}
Gapped phases of quantum matter are naturally described by topological quantum field theories (TQFT) at low energy and long distance.  For example, Abelian and non-Abelian Chern-Simons theories in $(2+1)d$ spacetime are believed to capture the topological properties of fractional quantum Hall states~\cite{wen-book, Nayak08}, and $\mathbb{Z}_2$ gauge theories(which can be described by $\mathrm{U}(1)\times \mathrm{U}(1)$ mutual Chern-Simons theory) have been proposed to describe some quantum spin liquids ($\mathbb{Z}_2$ spin liquids). Essentially these topological gauge theories encode nontrivial quantum statistics of low-energy excitations in a gapped phase. In two dimensions, low-energy excitations consist of localized quasiparticles, and their exchange and braiding statistics have been well understood.

In $(3+1)d$ spacetime, which is the dimension of the physical world, low-energy excitations are dramatically different: besides point-like particles, there can be loop-like excitations. A familar example is vortex lines in type-II superconductors. It is well-known that in three dimensions point-like particles can only have bosonic or fermionic exchange statistics, and no nontrivial mutual braiding statistics. On the other hand, there can be nontrivial braiding statistics between particles and loops, e.g. in discrete Abelian gauge theories.  Recently a new kind of braiding statistics between loop-like excitations, involving two loops linked to a third one, was discovered in Dijkgraaf-Witten gauge theories~\cite{Dijkgraaf90, Levin_PRL2014, Wang_arxiv2014, Ran_PRX2014, Wang_DW, Wan_DW, Moradi_arxiv2014, Jian_unpub, Bi_PRB2014, WW, CurtPRB2013, Curt_unpub, PutrovAP2017}.

Another impetus for interest in $(3+1)$-dimensional topological gauge theories comes from the study of symmetry-protected topological(SPT) phases~\cite{Chen_arxiv2011}. These are short-range entangled gapped phases, which in the absence of any symmetries are continuously connected to a trivial product state, but with certain symmetry $G$ they become topologically distinct. When $G$ is unitary, one can understand the SPT phases by promoting $G$ to a local gauge symmetry~\cite{LevinGu_arxiv2012}. Once the matter fields (i.e. SPT) are integrated out, one obtains a topological gauge theory at low energy. The nontriviality of SPT phases then manifests through the nontrivial braiding statistics of gauge fluxes in the gauged theory. This approach has been shown to correctly characterize all SPT phases with finite, Abelian unitary symmetries in both two and three dimensions~\cite{ChengPRL2014, Levin_PRL2014, Wang_arxiv2014,ChengPRX2018}.

Moreover, TQFTs also provide us a powerful tool to understand topological phase transitions. In fact, the continuum quantum fields in a TQFT should be regarded as emergent collective degrees of freedom in the vincinity of topological quantum phase transition, and the TQFT captures the topological Berry phase term induced by these collective degrees of freedom.  Given the important roles played by field theories in the study of two-dimensional topological phases and their phase transitions, it is desired to have a similar systematic construction of field theories in three dimensions.

In this paper we introduce TQFTs that describe nontrivial loop braiding statistics in $(3+1)d$ Abelian topological phases. The theories that we consider are all gauge theories, which naturally have non-local observables (Wilson loops and surfaces). Since we are interested in $(3+1)d$, the gauge theories involve both $1$-form and $2$-form gauge fields. We write down all possible Schwartz-type topological field theories that describe Abelian excitations. Namely, we require the action to be invariant under smooth diffeomorphisms, so they should be built out of the differential forms with wedge product. A similar approach was taken in \Ref{Wang_unpub} to write down response theories for SPT phases, where the gauge fields are treated as background fields. In contrast, we are interested in truly dynamical gauge theories.  An important point is that the action needs to have gauge invariance, in a manner whose precise meaning will be specified below. This requirement severely restricts possible terms that can appear in the action. We extract  particle and loop braiding statistics for some of these topological gauge theories which result in Abelian statistics. We hope this work will stimulate future theoretical studies on general $(3+1)d$ non-Abelian topological phases and topological phase transitions.

\section{General considerations}
Let us first discuss some general aspects of three-dimensional topologically ordered states.

\subsection{Excitations in 3D topological orders}

\begin{figure}
\centering
\includegraphics{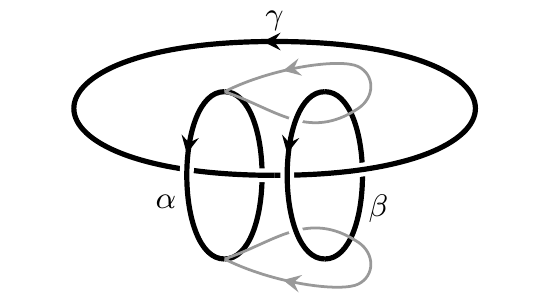}
\caption{Three-loop braiding process.}
\label{fig:threeloop}
\end{figure}

We list our physical assumptions of the general structures of topological excitations in 3D topological orders (TOs):
\begin{enumerate}
	\item 3D gapped topological phases can support two kinds of excitations: quasiparticles and quasi-strings. In the absence of boundary, quasi-strings always form closed loops.  We assume that there are a finite number of topologically distinct types of quasiparticles and quasi-strings\cite{tian2017}.
	\item For each type of quasi-string, one can create a single loop of this type out of the vacuum by a membrane operator. In other words, this single loop can be continuously shrinked to a local excitation. We say these are ``neutral'' loop excitations. On the other hand, if it shrinks to a topologically nontrivial quasi-particle, we say it is ``charged''.
	\item One can obviously define fusion of quasiparticles, as well as fusion of (unlinked) neutral loops.  Thus the set of quasiparticles form a unitary fusion category. In fact, they can further be endowed with braiding. However, because of the dimensionality, the braiding must be symmetric. This strongly constraints the structure of quasiparticles: the fusion category must be the category of irreducible linear representations of some finite group $G$, denoted by $\text{Rep}(G)$~\cite{Deligne}. They can have bosonic or fermionic exchange statistics. 
	\item There should be a generalized notion of braiding non-degeneracy in three dimensions. More concretely, there must be braiding processes that allow one to distinguish different types of quasiparticles from each other. Since braiding between quasiparticles are trivial, one has to use the braiding between quasiparticles and loops. In this regard, we only need unlinked single loops. It is then reasonable to postulate that one should be able to distinguish all types of quasiparticles by the braiding between quasiparticles and single neutral loops. Furthermore, such particle-loop braiding must be consistent with the fusion rules of quasiparticles: for a fixed type of loop excitation $\alpha$, denote the braiding between $\alpha$ and a quasiparticle of type $a$ by $B_{a,\alpha}$. Then
		\begin{equation}
			B_{a,\alpha}B_{b,\alpha}=\sum_{c}N_{ab}^c B_{c,\alpha}.
			\label{}
		\end{equation}
		Therefore, $B_{a,\alpha}$ defines a character on the $\text{Rep}(G)$ category. It is easy to see that such characters are nothing but the characters of the representations. Since characters are class functions, we have seen that each type of quasi-strings must correspond to a conjugacy class of $G$, uniquely. 
\item Braiding statistics between quasi-strings can be very complicated, since quasi-strings may be knotted and/or linked. It was proposed\cite{Levin_PRL2014, Ran_PRX2014} that the most fundamental braiding process of quasi-string braiding involves three loops (Fig.~\ref{fig:threeloop}): loop $\alpha$ is braided around loop $\beta$, while both are linked to a third loop $\gamma$. Simple two-loop process cannot capture the essence of 3D topological orders (TOs), and many complicated processes can be decomposed to a sequence of three-loop processes. So far, all known 3D TOs can be characterized by the three-loop braiding statistics. Nevertheless, whether the three-loop braiding statistics is complete for 3D TOs remains an open question.
\end{enumerate} 
Note that in this discussion, we assume both quasiparticles and quasistrings are free to move in space, and exclude the fracton topological order with immobile excitations~\cite{Haah, VijayPRB2015, VijayPRB2016}.

\subsection{Topological gauge theories in $(3+1)$d}
We aim to study topological gauge theories, with $1$-form and $2$-form gauge fields~\cite{KapustinArxiv2013}, to describe TOs in 3+1 dimensions. This is natural since $1$-form gauge fields minimally couple to worldlines of particles and $2$-form gauge fields couple to worldsheets of strings. We will only consider U$(1)$ gauge fields for simplicity.

To begin with, we enumerate all possible types of topological terms (dropping the indices for components of the gauge fields): 
\begin{align}
B\wedge\di A, \\
A\wedge A \wedge \di A,\\
A\wedge A\wedge A \wedge A,\\
B\wedge B,\\
B\wedge A \wedge A,
\end{align}
where $A$ is 1-form and $B$ is 2-form. $B\wedge \di A$ is the familiar BF term, describing the simplest discrete gauge theories. It is natural to include BF terms in the TQFT from the onset, for the following reason: to describe a discrete (i.e. $\mathbb{Z}_n$) gauge theory in a continuum formalism, we can start from a $\mathrm{U}(1)$ gauge field and add charge-$n$ bosonic matter fields. The Higgs phase effectively realizes a $\mathbb{Z}_n$ gauge theory. By performing a standard duality transformation, this Higgs theory can be rewritten as a topological BF theory. 

We now consider the other topological terms. Conventionally, we require that the Lagrangian is invariant (up to boundary terms) under the following gauge transformations:
\begin{equation}
	A\rightarrow A+\di f, B\rightarrow B+\di \xi.
	\label{}
\end{equation}
Here $f$ is a $\mathbb R$-valued function (mod $2\pi$) and $\xi$ is a $1$-form. This gauge-invariance condition then excludes all the other terms except the BF term. In order to describe more exotic statistical properties, it is necessary to generalize the notion of gauge transformations. For example, when gauge transforming $A$, we should also allow $B$ to transform:
\begin{equation}
	A\rightarrow A+\di f, B\rightarrow B + F[f, A].
	\label{eqn:strangegaugetrans1}
\end{equation}
Here $F[f,A]$ is a $2$-form built out of $f$ and $A$, such that $F[f=0, A]=0$. Similarly, when gauge transforming $B$ by $\di \xi$ we allow $A$ to be shifted by $\xi$:
\begin{equation}
	A\rightarrow A+c\xi, B\rightarrow B+\di \xi.
	\label{eqn:strangegaugetrans2}
\end{equation}
Here $c$ is constant.

Therefore, our first working assumption is that the allowed topological terms are those that can satisfy generalized gauge transformations Eq. \eqref{eqn:strangegaugetrans1} and Eq. \eqref{eqn:strangegaugetrans2} with appropriate choices of $F$ and $c$. This is different from the approach taken in \Ref{Wang_unpub}, where the topological terms are introduced as responses of SPT phases to non-dynamical symmetry gauge fields, and a flat connection condition is imposed to recover gauge invariance.

For simplicity, let us consider a BF theory with the other topological terms all of the same type in our list.  For all four types of topological terms, one can indeed find generalized gauge transformations to make sense of the topological gauge theories. We will focus on $A\wedge A \wedge \di A$ and $B\wedge B$ terms in the following sections. It has been proposed that $A\wedge A \wedge \di A$ type terms are responsible for three-loop braiding statistics~\cite{kapustin2014, Ye_unpub2015, Wang_unpub, Ryu2015, TiwariPRB2017}, and we will derive this result explicitly. We also found that $B\wedge B$ type terms can alter the exchange statistics of point-like excitations (i.e. from bosonic to fermionic). The SPT response theory indicates that the $A\wedge A\wedge A\wedge A$ type terms actually describe non-Abelian three-loop braiding statistics~\cite{Gu_unpub}. Recently, it has also been conjectured that the $B\wedge A\wedge A$ type terms are related to non-Abelian particle-loop braiding statistics\cite{chan17}.

\section{$A\wedge A\wedge \di A$ gauge theory}
Let us start with TQFT containing a cubic term $A_I A_J dA_K$. We will show that such TQFTs can describe the three-loop braiding statistics.

\subsection{A simple example}
To begin with, let us consider the following gauge theory with three gauge fields $A^I_\mu$ ($I=1,2,3$) corresponding to gauge group $G=\mathbb Z_{n_1} \times \mathbb Z_{n_2} \times \mathbb Z_{n_3}$:
\begin{align}\label{eq:L123}
\mathcal{L}=\sum_{I=1}^3\frac{n_I}{4\pi}\varepsilon^{\mu\nu\lambda\rho} B^I_{\mu\nu}\partial_\lambda A^I_{\rho} + \frac{M}{4\pi^2}\varepsilon^{\mu\nu\lambda\rho}A^1_{\mu}A^2_{\nu}\partial_\lambda A^3_{\rho}.
\end{align}
This Lagrangian is an example of the general theory \eq{eqn:dw} below with $M_{123}=2M$ and $M_{IJK}=0$ otherwise.

The Lagrangian \eq{eq:L123} is gauge invariant (up to total derivative) under the following gauge transformations:
\begin{align}\label{eq:GT}
\begin{aligned}
A_I&\rightarrow A_I+\di f_I, \quad (I=1,2,3),\\
B_1&\rightarrow B_1 +\di g_1 + \frac{M}{2\pi n_1}\di f_2 A_3,\\
B_2&\rightarrow B_2 +\di g_2 - \frac{M}{2\pi n_2}\di f_1 A_3,\\
B_3&\rightarrow B_3 +\di g_3.
\end{aligned}
\end{align}
Here, we have again defined $A_I=A^I_\mu \di x^\mu$ and $B_I=\frac{1}{2}B^I_{\mu\nu}\di x^\mu\wedge\di x^\nu$. And $f_I$ and $g_I$ ($I=1,2,3$) are 0-form and 1-form gauge transformation parameters respectively.

To quantize the theory \eq{eq:L123}, we first integrate over $B^I_{0\nu}$ and $A^I_{0}$, which impose the flat connection condition for $A^I_i$ and $B^I_{ij}$ on spacial manifold. Then we can use the standard canonical quantization procedure to quantize the theory \eq{eq:L123}, where only the terms with differential along time direction remain.

\subsubsection{Quantization and periodicity}
\label{sec:period}
 
Since $\di B_I$ is quantized as
\begin{align}
\int_\Omega \di B_I \in 2\pi\mathbb Z
\end{align}
for arbitrary closed surface $\Omega$ before and after the gauge transformation \eq{eq:GT}, we have the quantization of $M$: $M/n_1 \in \mathbb Z$ and $M/n_2 \in \mathbb Z$. Therefore the final quantization is $M \in n^{12}\mathbb Z$, where $n^{12}$ is the least common multiplier of $n_1$ and $n_2$.

It will be shown later that $M$ and $M+n^{12}n_{123}$ should be identified, for they give the same loop braiding statistics. Combined with the quantization of $M$ as an integer multiple of $n^{12}$, we see that $M=n^{12}p$ ($p=0,1,\cdots,n_{123}-1$).

\subsubsection{Membrane operators}
Observables in gauge theory are gauge invariant Wilson operators. In our theory, the gauge invariant Wilson loops are
\begin{align}
W_{I\gamma}&=\exp\left(i\Phi_{I\gamma}\right),\\
\Phi_{I\gamma}&=\oint_\gamma A_I,
\end{align}
where $\gamma$ is a closed curve. They are invariant under gauge transformation \eq{eq:GT}.

However, the usual Wilson surface operator $\exp(i\int_\Omega B_I)$ for a closed surface $\Omega$ in \emph{spacetime} is not gauge invariant. Therefore, we modify its definition to be
\begin{align}
	U_{I\Omega}&=\exp\left(iX_{I\Omega}\right),\\
	X_{1\Omega} &= \int_{\Omega} B_1 + \frac{M}{2\pi n_1} \int_V A_2\di A_3,\\\label{eqn:surfaceX2}
	X_{2\Omega} &= \int_{\Omega} B_2 - \frac{M}{2\pi n_2} \int_V A_1\di A_3,\\
	X_{3\Omega} &= \int_{\Omega} B_3,
\end{align}
where $V$ is a volume such that $\partial V=\Omega$. One can check that the new Wilson surface operators are indeed invariant under gauge transformation \eq{eq:GT}. In the canonical quantization, however, we only consider the Wilson surface operators in three dimensional \emph{spacial} manifold. Since $A_3$ is a flat connection in space in the canonical quantization procedure, we can drop the $A_I\di A_3$ term in the above definition and have a simpler expression for spacial Wilson surface operators:
\begin{align}
	U_{I\Omega}&=\exp\left(iX_{I\Omega}\right),\\\label{eqn:surfaceX1}
	X_{I\Omega} &= \int_{\Omega} B_I,
\end{align}
where $\Omega$ is a closed surface in three dimensional space.

\subsubsection{Canonical quantization}
We can do canonical quantization of the theory \eq{eq:L123}. 
By definition, the canonical momentum for $A^I_{i}$ is
\begin{align}
\pi^I_{i} &= \frac{n_I}{4\pi} \epsilon^{ijk} B^I_{jk},\quad (I=1,2)\\
\pi^3_{i} &= \frac{n_3}{4\pi} \epsilon^{ijk} B^3_{jk} + \frac{M}{4\pi^2} \epsilon^{ijk} A_{1j} A_{2k}.
\end{align}
Using the canonical quantization conditions for $A^I_i$ and $\pi^J_j$, one can show the commutation relations for $A$'s and $B$'s are
\begin{align}\label{eq:AA}\nonumber
[A^I_{i}(\mathbf x), A^J_{j}(\mathbf y)] &= 0,\\
[A^I_{i}(\mathbf x), B^J_{jk}(\mathbf y)] &= \frac{2\pi i}{n_I} \delta_{IJ} \epsilon_{ijk} \delta(\mathbf {x-y}),\\\nonumber
[B^1_{ij}(\mathbf x), B^3_{kl}(\mathbf y)] &= \frac{-i\delta(\mathbf{x-y})}{n_1 n_3} M \left[\epsilon^{lij} A^2_{k}(\mathbf x) - \epsilon^{ijk} A^2_{l}(\mathbf x) \right],\\\nonumber
[B^2_{ij}(\mathbf x), B^3_{kl}(\mathbf y)] &= \frac{i\delta(\mathbf{x-y})}{n_2 n_3} M \left[\epsilon^{lij} A^1_{k}(\mathbf x) - \epsilon^{ijk} A^1_{l}(\mathbf x) \right],\\\nonumber
[B^I_{ij}(\mathbf x), B^J_{kl}(\mathbf y)] &= 0,\quad (\text{other $I$ and $J$}).
\end{align}
The non-commutativity between $B$'s come from the requirement of $[\pi,\pi]=0$.

\subsubsection{Three-loop braiding}

The commutation relations of $A$'s and $B$'s contain the information of braiding statistics of point-like and loop-like excitations associated with the Wilson line and surface operators.

Consider a line $\gamma$ and a surface $\Omega$ intersecting transversely once. Using the commutation relation $[A,B]$, one can show that
\begin{equation}
	[ \Phi_{I\gamma}, X_{J\Omega}]=\delta_{IJ}\frac{2\pi i}{n_I}.
	\label{eq:XY}
\end{equation}
Therefore, we can obtain the group commutator of Wilson line and surface operators:
\begin{align}
K(W_{I\gamma}, U_{I\Omega}) &= {W_{I\gamma}}^\dagger {U_{J\Omega}}^\dagger W_{I\gamma} U_{J\Omega} \\\nonumber
&= e^{-[\Phi_{I\gamma},X_{J\Omega}]} = e^{-\frac{2\pi i}{n_I} \delta_{IJ} }
\end{align}
This is the well-known result that braiding a species $I$ charge around a fundamental flux line of $\mathbb Z_{n_J}$ gauge theory of species $J$ gives a statistics phase $\frac{2\pi i}{n_I} \delta_{IJ}$.
In the following, we will show there is non-trivial three-loop braiding statistics for the theory \eq{eq:L123}.

The Berry phase accumulated in the process corresponds to the three-loop braiding, where two loops with unit $I$ and $J$ fluxes are linked to a base loop with unit $K$ flux, can be calculated as \cite{Levin_PRL2014,Ran_PRX2014,Yoshida} 
\begin{align}\label{eq:3loop123}
e^{i\theta_{IJ,K}}
&= K(K(U_{Iyz}, U_{Jzx}),U_{Kxy})\\\nonumber
&= \exp\left( {-i\left[\left[X_{Iyz},X_{Jzx}\right],X_{Kxy}\right]} \right),
\end{align}
where we have chosen the three surfaces to be $xy, yz$ and $zx$ plane. The basic idea of the process is that $U_{Kxy}$ first create a base loop $K$, and $K(U_{Iyz}, U_{Jzx})$ do a full braiding of two other loops linked to $K$, then annihilate the base loop and do a full braiding of the two other loops. Using commutation relations of $A$'s and $B$'s, one can directly show that the only non-trivial three-loop braiding phase factors are
\begin{align}\label{eq:theta123}
\begin{aligned}
e^{i\theta_{13,2}} &= e^{i\theta_{31,2}} = \exp\left( -\frac{2\pi M i}{n_1 n_2 n_3} \right),\\
e^{i\theta_{23,1}} &= e^{i\theta_{32,1}} = \exp\left( \frac{2\pi M i}{n_1 n_2 n_3} \right).
\end{aligned}
\end{align}
The topological invariant for the three-loop braiding is $e^{i\Theta_{IJ,K}}=e^{in^{IJ}\theta_{IJ,K}}$ \cite{Wang_arxiv2014}. So the nontrivial ones are
\begin{align}
\begin{aligned}
e^{i\Theta_{13,2}} &= e^{i\Theta_{31,2}} = \exp\left( -\frac{2\pi n^{13} M i}{n_1 n_2 n_3} \right),\\
e^{i\Theta_{23,1}} &= e^{i\Theta_{32,1}} = \exp\left( \frac{2\pi n^{23} M i}{n_1 n_2 n_3} \right).
\end{aligned}
\end{align}
From the above expressions, we see that $M$ and $M+n^{12}n_{123}$ give the same topological invariants. Therefore we identify these two values of $M$ and have $M=n^{12}p$ ($p=0,1,\cdots,n_{123}-1$).

Apart from the three-loop braiding calculated above, there are also processes of three-loop half-braidings when two loops $I=J$ are linked to $K$. We can first create a base loop by $U_{Kxy}$. Since the full braiding process is given by $K(U_{Iyz}, U_{Jzx})$ in \eq{eq:3loop123}, we can do a half-braiding by $U_{Iyz} U_{Izx}$, which do not move the two loops back to its original places. Therefore, the three-loop half-braiding phase can be calculated by
\begin{align}\label{eq:thetaIK}
e^{i\theta_{I,K}}
&= \langle 0| K(U_{Iyz} U_{Izx},U_{Kxy}) |0\rangle.
\end{align}
Using the canonical commutation relations and the fact $|0\rangle$ is a state without any flux loops ($\langle 0|\Phi_{I\gamma}|0\rangle$=0), one can show directly that all half-braiding phases $\theta_{I,K}$ are $0$ for this particular theory \eq{eq:L123}.

\subsection{General theory}

Now let us consider the most general partition function of the $A\wedge A\wedge dA$ type TQFT:
\begin{align}\label{eq:Z}
Z = \int \mathcal D B \mathcal DA e^{i\int\di^4 x \mathcal L[B,A]},
\end{align}
where the Lagrangian is given by
\begin{equation}
	\mathcal{L}=\frac{n_I}{4\pi}\varepsilon^{\mu\nu\lambda\rho} B^I_{\mu\nu}\partial_\lambda A^I_{\rho} + \frac{M_{IJK}}{8\pi^2}\varepsilon^{\mu\nu\lambda\rho}A^I_{\mu}A^J_{\nu}\partial_\lambda A^K_{\rho}.
	\label{eqn:dw}
\end{equation}
Here all the repeated indexes are summed over automatically. Although we can choose $M_{IJK}+M_{JIK}=0$ without loss of generality, we would \emph{not} to impose this condition for the coefficients in the following discussions.

Naively the theory is not invariant under the gauge transformation of $A_I=A^I_{\mu}\di x^\mu$. To recover gauge invariance, we need to let the gauge transformation also acts on $B_I=\frac{1}{2}B^I_{\mu\nu}\di x^\mu \wedge\di x^\nu$:
\begin{align}
\begin{aligned}
		A_I&\rightarrow A_I+\di f_I,\\
		B_I&\rightarrow B_I +\di g_I + \frac{M_{IJK}-M_{JIK}}{4\pi n_I}\di f_J A_K,
	\label{eqn:gaugetrans1}
\end{aligned}
\end{align}
where the gauge parameters are quantized as $\int_\gamma\di f_i\in 2\pi\mathbb Z$ and $\int_\Omega\di g_i\in 2\pi\mathbb Z$ on closed line $\gamma$ and surface $\Omega$.
It is easy to check that the theory is indeed gauge-invariant with this definition of gauge transformations. We notice that one can come up with different gauge transformations to make the action gauge-invariant, however our choice in Eq. \eqref{eqn:gaugetrans1} is motivated and justified by a microscopic derivation of the action \eqref{eqn:dw} (with non-compact $B$ fields). We note that the gauge transformation define here is different from the one defined in Ref.~\onlinecite{Ryu2015}. In Appendix B, we will provide microscopic derivation of such a twisted gauge transformation.

\subsubsection{Quantization and Periodicity}

Quantization of $B$ requires $(M_{IJK}-M_{JIK})/(2n_I)\in\mathbb{Z}$ and $(M_{JIK}-M_{IJK})/(2n_J)\in\mathbb{Z}$, so $(M_{IJK}-M_{JIK})/2$ must be an integer multiple of $n^{IJ}$. 
We will show later that the theories $(M_{IJK}-M_{JIK})/2$ and $(M_{IJK}-M_{JIK})/2+n^{IJ}n_{IJK}$ have the same braiding invariants. Combine this identification with the previous quantization, we have $(M_{IJK}-M_{JIK})/2=n^{IJ}p$ ($p=0,1,\cdots,n_{IJK}-1$).

\subsubsection{Membrane Operators}
 We now compute the physical observables in the theory. Due to the cubic form of \eqref{eqn:dw}, we are no longer able to integrate out the gauge fields exactly to obtain an effective action of matter fields. Therefore we proceed with canonical quantization.

First let us define gauge-invariant physical observables. Wilson loops take the conventional form
\begin{equation}
	W_{I\gamma}=\exp\left(i\oint_\gamma A_I\right).
	\label{}
\end{equation}
Here $\gamma$ is any closed curve. For later use, we will also define
\begin{equation}
	\Phi_{I\gamma}=\oint_\gamma A_I.
\end{equation}
The gauge invariant Wilson surface operators for a closed surface $\Omega$ in spacetime are
\begin{align}\label{full_transf}
U_{I\Omega}&=\exp\left(iX_{I\Omega}\right),\\
X_{I\Omega} &= \int_{\Omega} B_I + \frac{M_{IJK}}{2\pi n_I} \int_V A_J\di A_K,
\end{align}
where $V$ is a volume such that $\partial V=\Omega$. One can check that the Wilson surface operators, with an additional Chern-Simons density term compared to the usual definition, are invariant under the gauge transformation \eq{eqn:gaugetrans1}. In the canonical quantization, we only need to consider the Wilson surface operators in three dimensional \emph{spacial} manifold. After integrating out the Lagrangian multiples $B^I_{0i}$ and obtaining the flat connections $A_I$ in space, we can drop the $A_J\di A_K$ terms in the above definition and have a simpler expression for spacial Wilson surface operators:
\begin{align}
	U_{I\Omega}&=\exp\left(iX_{I\Omega}\right),\\\label{eqn:surfaceX}
	X_{I\Omega} &= \int_{\Omega} B_I.
\end{align}

\subsubsection{Canonical quantization and Membrane Algebra}

To carry out canonical quantization, the four manifold has to be $\mathcal{M}=\mathcal{M}_3\times\mathbb{R}$ where $\mathbb{R}$ corresponds to the time direction. Again, the time components $A_{I0}, B_{I0i}$ are all Lagrange multipliers and just enforce the constraint that $\varepsilon^{ij}\partial_i A_{Ij}=0, \varepsilon^{ijk}\partial_i B_{Ijk}=0$ when there are no external sources,
and the Hilbert spaces are flat connections of $A$ and $B$ modulo gauge transformations. 

By definition, the canonical momentum for $A_{Ii}$ is
\begin{align}\label{eq:pi}
\pi_{Ii} = \frac{\partial\mathcal L}{\partial(\partial_0 A_{Ii})} = \frac{n_I}{4\pi} \epsilon^{ijk} B_{Ijk} + \frac{M_{JKI}}{8\pi^2} \epsilon^{ijk} A_{Jj} A_{Kk}.
\end{align}
From the canonical quantization conditions
\begin{subequations}
\begin{align}
[A_{Ii}(\mathbf x), A_{Jj}(\mathbf y)] &= 0,\\
[A_{Ii}(\mathbf x), \pi_{Jj}(\mathbf y)] &= i \delta_{IJ} \delta_{ij} \delta(\mathbf {x-y}),\\
[\pi_{Ii}(\mathbf x), \pi_{Jj}(\mathbf y)] &= 0,
\end{align}
\end{subequations}
we obtain the commutation relations between fields $A$'s and $B$'s:
\begin{widetext}
\begin{subequations}
\begin{align}
[A_{Ii}(\mathbf x), A_{Jj}(\mathbf y)] &= 0,\\\label{eq:AB}
[A_{Ii}(\mathbf x), B_{Jjk}(\mathbf y)] &= \frac{2\pi i}{n_I} \delta_{IJ} \epsilon^{ijk} \delta(\mathbf {x-y}),\\\label{eq:BB}
[B_{Iij}(\mathbf x), B_{Jkl}(\mathbf y)] &= \sum_K \frac{i\delta(\mathbf{x-y})}{n_I n_J} \left[ M_{IKJ}\epsilon^{ijk} A_{Kl}(\mathbf x) + M_{KIJ}\epsilon^{ijl} A_{Kk}(\mathbf x) - M_{JKI}\epsilon^{ikl} A_{Kj}(\mathbf x) - M_{KJI}\epsilon^{jkl} A_{Ki}(\mathbf x) \right].
\end{align}
\end{subequations}
If we consider the commutation relation $[\frac{1}{2}(B_{Iij}-B_{Iji})(\mathbf x),\frac{1}{2}(B_{Jkl}-B_{Jlk})(\mathbf y)]$, we need only to antisymmetrize the indices $i$ and $j$, $k$ and $l$ for the last equation:
\begin{align}\label{eq:BB2}
&\quad\left[\frac{1}{2}(B_{Iij}-B_{Iji})(\mathbf x),\frac{1}{2}(B_{Jkl}-B_{Jlk})(\mathbf y)\right]\\\nonumber
&=
 \sum_K \frac{i\delta(\mathbf{x-y})}{n_I n_J} \left[ \frac{M_{IKJ}-M_{KIJ}}{2} \left(\epsilon^{ijk} A_{Kl}(\mathbf x) -\epsilon^{ijl} A_{Kk}(\mathbf x) \right) + \frac{M_{KJI}-M_{JKI}}{2} \left(\epsilon^{ikl} A_{Kj}(\mathbf x) - \epsilon^{jkl} A_{Ki}(\mathbf x) \right) \right].
\end{align}
\end{widetext}
This equation turns out to be related to the three-loop braiding statistics. We also note that only the antisymmetric part of the first two indices of $M_{IJK}$ appears in the above commutation relation.

For closed line $\gamma$ and closed surface $\Omega$ intersecting transversely, we have the commutation relation between the Wilson loop and surface operators:
\begin{equation}
	[ \Phi_{I\gamma}, X_{J\Omega}]=\delta_{IJ}\frac{2\pi i}{n_I}I(\gamma, \Omega).
	\label{eq:XY}
\end{equation}
Here
\begin{align}
I(\gamma, \Omega) = \sum_{p\in \gamma\cap \Omega} \operatorname{sgn}(p) = \sum_{p\in \gamma\cap \Omega} \operatorname{sgn}\left[\hat n_\gamma(p) \cdot \hat n_\Omega(p)\right],
\end{align}
is the signed intersection number of $\gamma$ and $\Omega$. $\hat n_\gamma(p)$ [$n_\Omega(p)$] is the tangent (normal) direction of $\gamma$ ($\Omega$) at point $p$. If $\gamma$ and $\Omega$ intersect non-transversely, then the commutation relation is zero due to Eq.~(\ref{eq:AB}).

By using the Baker-Campbell-Hausdorff formula, we can obtain the group commutator of Wilson line and surface operators:
\begin{align}
K(W_{I\gamma}, U_{I\Omega}) &= {W_{I\gamma}}^\dagger {U_{J\Omega}}^\dagger W_{I\gamma} U_{J\Omega} \\\nonumber
&= e^{-[\Phi_{I\gamma},X_{J\Omega}]} = e^{-\frac{2\pi i}{n_I} \delta_{IJ} I(\gamma,\Omega)}
\end{align}
This is the well-known result that braiding a species $I$ charge around a fundamental flux line of $\mathbb Z_{n_J}$ gauge theory of species $J$ gives a statistics phase $\frac{2\pi i}{n_I} \delta_{IJ}$.

\subsubsection{Three loop braiding}

Now let us move to the statistics of loops. We assume two closed surfaces $\Omega$ and $\Omega'$ embedded in $\mathcal{M}_3$ intersect transversely. The intersection $\Omega\cap \Omega'$ is then a collection of (directed) closed lines. The direction of the line is given by $\hat n_\Omega\times \hat n_{\Omega'}$ locally, where $\hat n_\Omega$ is the local normal direction of the surface $\Omega$. Using the convention $B_I=\frac{1}{2}B^I_{\mu\nu}\di x^\mu\wedge\di x^\nu$ and the commutation relations \eq{eq:BB2}, one can show straightforwardly that
\begin{align}\label{eq:YY}
&\quad[X_{I\Omega},X_{J\Omega'}] \\\nonumber
&= \sum_K \frac{i}{n_I n_J} \frac{M_{KIJ}-M_{IKJ}+M_{KJI}-M_{JKI}}{2} \Phi_{K,\Omega\cap \Omega'}.
\end{align}
The line integral in $\Phi_{K,\Omega\cap \Omega'}$ on the right hand side is along the direction defined above.

Using Eqs. (\ref{eq:XY}) and (\ref{eq:YY}), one can further show that
\begin{align}
&\quad\left[\left[X_{I\Omega},X_{J\Omega'}\right],X_{K\Omega''}\right] \\\nonumber
&= \frac{\pi}{n_I n_J n_K} (M_{IKJ}-M_{KIJ}+M_{JKI}-M_{KJI}) I(\Omega\cap \Omega', \Omega'').
\end{align}
Note that $\Omega\cap \Omega'$ does not intersect transversely with neither $\Omega$ or $\Omega'$. Therefore, $\Phi_{K,\Omega\cap \Omega'}$ commutes with both $X_{I\Omega}$ and $X_{J\Omega'}$. And by using the Baker-Campbell-Hausdorff formula again, we have
\begin{align}
K(U_{I\Omega}, U_{J\Omega'})
&= {U_{I\Omega}}^\dagger {U_{J\Omega'}}^\dagger U_{I\Omega} U_{J\Omega'}
=e^{-[X_{I\Omega},X_{J\Omega'}]},
\end{align}
and finally
\begin{align}
&\quad\ K(K(U_{I\Omega}, U_{J\Omega'}),U_{K\Omega''})\\\nonumber
&= \exp\left( {-i\left[\left[X_{I\Omega},X_{J\Omega'}\right],X_{K\Omega''}\right]} \right)\\\nonumber
&= \exp \bigg[-\frac{\pi i}{n_I n_J n_K} (M_{IKJ}-M_{KIJ}+M_{JKI}-M_{KJI})\\\nonumber
&\quad\cdot I(\Omega\cap \Omega', \Omega'') \bigg]
\end{align}

In order to reveal the nontrivial statistical properties, we consider $\mathcal{M}_3={T}^3$ whose dimensions we refer to as $x,y,z$. We will use $i,j,k$ to denote the three spatial directions. 
The nontrivial braiding statistics of particle and loop excitations manifests in the algebra of Wilson operators defined on non-contractible cycles and surfaces (i.e. nontrivial cohomology classes in $H^1(\mathcal{M}_3, \mathbb{Z})$ and $H^2(\mathcal{M}_3, \mathbb{Z})$). By definition, such surfaces are not the boundary of any open volume. As illustrated above, we can use alternative definitions of Wilson surface operators purely on the surface. And the commutation relations, hence the braiding statistics, are all the same for these operators.
For $\mathcal{M}_3=T^3$, we denote the non-contractible cycles by $x,y,z$ and the three nontrivial surfaces by $xy, yz, zx$.

According to Ref.~\onlinecite{Levin_PRL2014,Ran_PRX2014,Yoshida}, the Berry phase accumulated in the process corresponds to the three-loop braiding, where two loops with unit $I$ and $J$ fluxes are linked to a base loop with unit $K$ flux, is
\begin{align}\label{eq:3loop}
e^{i\theta_{IJ,K}}
&= K(K(U_{Iyz}, U_{Jzx}),U_{Kxy})\\\nonumber
&= \exp\left[ {\frac{-\pi i}{n_I n_J n_K} (M_{IKJ}-M_{KIJ}+M_{JKI}-M_{KJI}) }\right],
\end{align}
where we have used the fact $I(\Omega_{yz}\cap\Omega_{zx},\Omega_{xy})=1$.

Since $(M_{IJK}-M_{JIK})/2$ is an integer multiple of $n^{IJ}$, we can parametrize it to be $M_{IJK}-M_{JIK}=2n^{IJ}m_{IJK}$ where $m_{IJK}$ is an integer and $n^{IJ}$ is the least common multiple of $n_I$ and $n_J$. The topological invariant for the three-loop braiding is then given by
\begin{align}\label{eq:res}
&\quad e^{i\Theta_{IJ,K}} = e^{in^{IJ}\theta_{IJ,K}}\\\nonumber
&= \exp\left[ -i\left(\frac{2\pi n^{IJ}}{n_{IK} n_J} m_{IKJ} + \frac{2\pi n^{IJ}}{n_{JK} n_I} m_{JKI} \right)\right].
\end{align}
This is consistent with the results in Ref.~\onlinecite{Wang_arxiv2014}.

Similar to the three-loop braiding phase factor $e^{i\theta_{IJ,K}}$, the three-loop half-braiding phase factor $e^{i\theta_{I,K}}$ can also be obtained. Naively, we have $\theta_{I,K}=\theta_{II,K}/2$, since twice the half-braiding is a full-braiding. But there is an ambiguity of $\pi$ in $\theta_{II,K}/2$, for both $\theta_{II,K}$ and $\theta_{I,K}$ are defined modulo $2\pi$.

Let us calculate in detail the three-loop half-braiding statistics from the canonical quantization. Similar to the expression \eq{eq:3loop}, the process of three-loop half-braiding can be written as (see the discussions above \eq{eq:thetaIK})
\begin{align}
e^{i\theta_{I,K}}
&= K(U_{Iyz} U_{Izx},U_{Kxy}).
\end{align}
Using the Baker-Campbell-Hausdorff formula, we have
\begin{align}\label{eq:UU}
U_{Iyz} U_{Izx} = \exp\left( iX_{Iyz} + iX_{Izx} - \frac{1}{2}[X_{Iyz},X_{Izx}] - i c \right),
\end{align}
where $c=\frac{1}{12}([X_{Iyz},[X_{Iyz},X_{Izx}]]+[X_{Izx},[X_{Izx},X_{Iyz}]])$ is a real number by using the commutation relations of $X$'s and $\Phi$'s. The three-loop half-braiding phase factor is then
\begin{align}
e^{i\theta_{I,K}}
&= \langle 0| K(U_{Iyz} U_{Izx},U_{Kxy}) |0\rangle\\\nonumber
&= \langle 0| K\left(\exp\left(- \frac{1}{2}[X_{Iyz},X_{Izx}]\right),U_{Kxy}\right) |0\rangle\\\nonumber
&= \exp\left[ {\frac{-\pi i}{n_I^2 n_K} (M_{IKI}-M_{KII}) }\right],
\end{align}
which is exactly the naive result: half of the full-braiding result \eq{eq:3loop} or \eq{eq:res}. Note that the term $iX_{Iyz}+iX_{Izx}$ in \eq{eq:UU} contributes to the half-braiding phase factor as terms like $\langle 0| e^{i\Phi} |0 \rangle$, which is trivial for $|0\rangle$ is the ground state without any flux, i.e., $\Phi_{I\gamma}|0\rangle=0$.

From these braiding invariants, we see that $(M_{IJK}-M_{JIK})/2$ and $(M_{IJK}-M_{JIK})/2+n^{IJ}n_{IJK}$ would give the same results. Combined with the quantization of $M$, we have the result $(M_{IJK}-M_{JIK})/2=n^{IJ}p$ ($p=0,1,\cdots,n_{IJK}-1$) claimed above.

\section{$B\wedge B$-type gauge theory}
The other Abelian family of TQFT is given by the following action:
\begin{equation}
	\mathcal{L}= \frac{n_I}{4\pi}\varepsilon^{\mu\nu\lambda\sigma}B_{I\mu\nu} \partial_\lambda A_{I\sigma}+\frac{K_{IJ}}{16\pi} \varepsilon^{\mu\nu\lambda\sigma}B_{I\mu\nu}B_{J\rho\sigma}.
	\label{eqn:WWaction2}
\end{equation}
The single-component version of the gauge theory \eqref{eqn:WWaction2} was first introduced in \Ref{HorowitzCMP}, and its relevance to 3D topological phases of matter has been increasingly appreciated in recent years~\cite{WW, CurtPRB2013, Curt_unpub, Ye_unpub}. 
Naively, the action is not invariant under the gauge transformations of $B$. As observed in \Ref{HorowitzCMP}, gauge invariance can be achieved with the following generalized gauge transformations 
\begin{equation}
	\begin{aligned}
		B_{I\mu\nu}&\rightarrow B_{I\mu\nu}+\partial_{[\mu}\xi_{I\nu]}\\
		A_{I\mu}&\rightarrow A_{I\mu}-\frac{K_{IJ}}{n_I}\xi_{J\mu}
	\end{aligned}
	\label{eqn:1formgauge}
\end{equation}
Notice that with this definition, $F_{\mu\nu}=\partial_{[\mu}A_{I\nu]}$ is no longer gauge-invariant. Instead, $G_{I\mu\nu}=F_{I\mu\nu}+\frac{K_{IJ}}{n_I}B_{J\mu\nu}$ can be used to construct a Maxwell-type kinetic term (in addition to the Maxwell term for $B$ built from the $3$-form curvature tensor of $B$). The above $B \wedge F+B \wedge B$ type TQFT can be formally rewritten as a $G \wedge G$ type term.(We note that the $F\wedge F$ type term is a total derivative which is dropped away here.) 


Let us check the gauge invariance explicitly, which will also lead to a quantization condition for $K$. For now let us assume that the theory is defined on a closed $4$-manifold. The variation of the action under the $2$-form gauge transformation becomes
\begin{equation}
	\delta \mathcal{S}=\frac{n_I}{4\pi}\varepsilon^{\mu\nu\lambda\sigma}\partial_{[\mu}\xi_{I\nu]}\partial_\lambda A_{I\sigma}-\frac{K_{IJ}}{16\pi}\varepsilon^{\mu\nu\lambda\sigma}\partial_{[\mu}\xi_{I\nu]}\partial_{[\rho}\xi_{J\sigma]}
	\label{}
\end{equation}
The first term is a total derivative. Integral of the second term is quantized:
\footnote{This is the intersection form on the $4$-manifold.}
\begin{equation}
	\frac{1}{16\pi^2}\int_\mathcal{M} \di^4x\,\varepsilon^{\mu\nu\lambda\sigma}\partial_{[\mu}\xi_{I\nu]}\partial_{[\lambda}\xi_{J\sigma]}\in \mathbb{Z}.
	\label{eqn:intersection}
\end{equation}
For the action to be gauge-invariant on any space-time manifold, the second term must an integral multiple of $2\pi $ which requires $K_{II}$ and $2K_{IJ}, I\neq J$ to be even integers.  Therefore we find a quantization condition 
\begin{equation}
	K_{II}\in 2\mathbb{Z}, K_{IJ}\in \mathbb{Z}.
	\label{}
\end{equation}
Notice, however, that on a spin manifold, the \eqref{eqn:intersection} quantizes to an even integer. So $K_{IJ}$ can be any integer if we are considering fermionic theories which can only be defined on spin manifolds. We will see that if any of $K_{II}$ is odd, the theory indeed admits transparent fermionic excitations.

On the other hand, we notice that because $A$ is compact, in order to keep the $2\pi$ periodicity in \eqref{eqn:1formgauge}, $\frac{K_{IJ}}{n_I}$ should be an integer. Similarly, $\frac{K_{JI}}{n_J}$ is also an integer. So $K_{IJ}$ is a multiple of $\mathrm{lcm}(n_I, n_J)$ where lcm means the least common multiple. We will write
\begin{equation}
	K_{IJ}=\mathrm{lcm}(n_I, n_J)k_{IJ}.
	\label{}
\end{equation}
The only constraint then is that $k_{II}n_I$ is even.

\subsection{Observables}
We now compute the physical observables in the quantum theory. To motivate,
let us couple the gauge fields to sources:
\begin{equation}
	\mathcal{L}_\text{coupling}=j_I^\mu A_{I\mu}+\frac{1}{2}\Sigma_{I}^{\mu\nu}B_{I\mu\nu}.
	\label{}
\end{equation}
First we need to make sure that the coupling term is gauge-invariant. Invariance under $1$-form gauge transformation gives the usual conservation law:
	$\partial_\mu j^\mu_I=0$. 
However, under the $2$-form gauge transformation
\begin{equation}
	\delta \mathcal{L}_\text{coupling}=-\frac{K_{IJ}}{n_I}\xi_{J\mu}j_I^\mu+\frac{1}{2}\Sigma_I^{\mu\nu}(\partial_\mu\xi_{I\nu}-\partial_\nu\xi_{I\mu}).
	\label{}
\end{equation}
So we must impose a different conservation law 
\begin{equation}
	\frac{K_{IJ}}{n_J}j_J^\mu+\partial_\nu\Sigma^{\mu\nu}_I=0.
	\label{eqn:conservationlaw}
\end{equation}
The physical interpretation is that point-like excitations are the end of string-like excitations. If $K_{IJ}=0$, all strings are closed loops.

	A completely equivalent viewpoint is to consider the expectation values of gauge-invariant operators, which are Wilson loops and surfaces. The Wilson surface operators are defined as
	\begin{equation}
		W_{I\Omega}=\exp\left(i\int_\Omega B_I\right).
		\label{}
	\end{equation}
Here $\Omega$ is a closed surface. One might attempt to construct Wilson loop operators as
\begin{equation}
	W_{I\gamma}=\exp\left( i\oint_\gamma A_I\right).
	\label{}
\end{equation}
However, it is no invariant under $2$-form gauge transformations. In order to restore gauge invariance, we have to attach to $\gamma$ a surface $\Omega$ such that $\partial \Omega=\gamma$ and define
\begin{equation}
	W_{I\gamma}=\exp\left(i\oint_\gamma A_I+i\frac{K_{IJ}}{n_J}\int_\Sigma B_J\right).
	\label{}
\end{equation}
It is easy to see that evaluating the expectation values of Wilson loop/surface operators using path integrals is the same as computing the path integral in the presence of sources.

We can integrate out the gauge fields to obtain an effective action of the source fields. Since the action is Gaussian, let us write down the equations of motion first:
\begin{equation}
	\begin{gathered}
	j_I^\mu+\frac{n_I}{4\pi}\varepsilon^{\mu\nu\lambda\sigma}\partial_\nu B_{I\lambda\sigma}=0\\
	\Sigma_I^{\mu\nu}+\frac{n_I}{2\pi}\varepsilon^{\mu\nu\lambda\sigma}\partial_\lambda A_{I\sigma}+\frac{K_{IJ}}{4\pi}\varepsilon^{\mu\nu\lambda\sigma}B_{J\lambda\sigma}=0\\
	\end{gathered}
	\label{}
\end{equation}
In the Lorentz gauge $\partial^\mu B_{I\mu\nu}=0$, we find
\begin{equation}
	\begin{gathered}
	B_{I\mu\nu}=-\frac{4\pi}{n_I}\varepsilon_{\mu\nu\lambda\rho}\frac{\partial^\lambda}{\square}j_I^\rho\\
-\frac{n_I}{2\pi}\varepsilon^{\mu\nu\lambda\sigma}\partial_\lambda A_{I\sigma}=-\Sigma_I^{\mu\nu}-\frac{K_{IJ}}{4\pi}\varepsilon^{\mu\nu\lambda\sigma}B_{J\lambda\sigma},
	\end{gathered}
	\label{eqn:eofsol}
\end{equation}
where $\square\equiv\partial^\mu\partial_\mu$. Substituting \eqref{eqn:eofsol} into \eqref{eqn:WWaction2}, we obtain an effective action:
\begin{equation}
	\begin{split}
		\Gamma[j,\Sigma]=&-\underbrace{\frac{2\pi}{n_I}\int_\mathcal{M}\di^4x\,\varepsilon_{\mu\nu\lambda\rho}\Sigma_I^{\mu\nu}\frac{\partial^\lambda}{\square}j_I^\rho}_{\Gamma_1}\\
		&-\underbrace{\frac{4\pi K_{IJ}}{n_In_J}\int_\mathcal{M}\di^4x\, \varepsilon_{\mu\nu\lambda\rho}\frac{\partial^\mu}{\square}j_I^\nu \frac{\partial^\lambda}{\square}j_I^\rho}_{\Gamma_2}
	\end{split}
	\label{eqn:S2}
\end{equation}

Particle current is defined by the worldlines:
\begin{equation}
	j_I^\mu(x)=\sum_aq_{Ia}\oint_{\gamma_a}\di y_{a}^\mu\,\delta(x-y_{a}).
	\label{}
\end{equation}
 $\gamma_a$ is the worldline of the particle $a$ carrying $q_{Ia}\in \mathbb{Z}$ charges of the gauge field $A_I$. 
 
 The string current needs some care~\cite{LechnerJHEP2000}. As we have noted, the gauge structure of the theory requires that each worldline bounds a (open) worldsheet, the choice of which is not unique. We will use the following worldsheet:
\begin{equation}
	\Sigma_a^{\mu\nu}(x)=u^\mu\int_0^\infty\di s\oint_{\gamma_a}\di y_a^\nu\,\delta(x-y_a-su).
	\label{}
\end{equation}
Here $u^\mu$ is a constant $4$-vector satisfying $u^2\neq 0$. It is straightforward to see that $\partial_\mu \Sigma_a^{\mu\nu}=j_a^\nu$. Besides $\Sigma_a$, there are closed worldsheet current $\Sigma'_I$ corresponding to the motion of flux loops:
\begin{equation}
	{\Sigma'}_I^{\mu\nu}(x)=\sum_{b}\phi_{Ib}\int_{\Omega_b}\di^4\sigma^{\mu\nu}(X_b)\delta(x-X_b).
	\label{}
\end{equation}
$X_b$ is the embedding of the worldsheet $\Omega_b$ into $\mathcal{M}$. The total worldsheet current is given by 
\begin{equation}
	\Sigma_I=\sum_a\frac{K_{IJ} q_{Ja}}{n_J}\Sigma_a + \Sigma'_I.
	\label{}
\end{equation}

\comments{
\begin{equation}
	\begin{split}
	\partial_\mu \Sigma_\gamma^{\mu\nu}(x)&=u^\mu\int\di\tau\,\int_0^\infty\di s\,\frac{\di y^\nu(\tau)}{\di\tau}\frac{\partial}{\partial x^\nu}\delta(x-y(\tau)-su)\\
&=u^\mu\int\di\tau\,\int_0^\infty\di s\,\frac{\di y^\nu(\tau)}{\di\tau}\frac{\partial}{\partial x^\mu}\delta(x-y(\tau)-su)\\
&=-\int\di\tau\,\int_0^\infty\di s\,\frac{\di y^\nu(\tau)}{\di\tau}\frac{\partial}{\partial s}\delta(x^\mu-y^\mu(\tau)-su^\mu)
	\end{split}
	\label{}
\end{equation}
}

We can then evaluate the effective action. We skip the details, which can be found in Appendix \ref{sec:bb}.  We find that the effective action is given by
\begin{equation}
	\begin{split}
	\Gamma[j,\Sigma]=&-\frac{\pi K_{IJ}}{n_In_J}\Big[\sum_{a\neq b}q_{Ia}q_{Jb}I(\gamma_a,\gamma_b)+\sum_a{q_{Ia}q_{Ja}}I(\gamma_a)\Big]\\
	&-\frac{2\pi q_{Ia}\phi_{Ib}}{n_I}I(\gamma_a,\Omega_b).
	\end{split}
	\label{eqn:effectiveactionfinal}
\end{equation}
where $I(\gamma_a,\gamma_b)$ for $a\neq b$ is the linking number of the 3D projection of $\gamma_a$ and $\gamma_b$:
\begin{equation}
	I(\gamma_a,\gamma_b)=\frac{1}{4\pi}\varepsilon_{ijk}\oint_{\gamma_a}\di y_a^i\, \oint_{\gamma_b}\di y_b^j\, \frac{y_a^k-y_b^k}{|\mb{y}_a-\mb{y}_b|^3}.
	\label{}
\end{equation}
$I(\gamma_a)$ is the self-linking number of the 3D projection of $\gamma_a$, which can be thought as the regularized version of $I(\gamma_a,\gamma_a)$.

As expected, the effective action is purely topological. However, the result does not make sense at first glance: the first two term in $\Gamma$ mean that particles can have nontrivial mutual braiding statistics and exchange statistics other than fermionic and bosonic ones, which is impossible in 3D. In fact, a ``particle'' labeled by a charge vector $\mb{q}=(q_1, q_2, \dots)$ has an exchange statistics 
\begin{equation}
	\theta_\mb{q}=e^{\frac{i\pi K_{IJ}q_Iq_J}{n_In_J}}.
	\label{}
\end{equation}
And two particles with charge vectors $\mb{q}$ and $\mb{q}'$ has mutual braiding statistics
\begin{equation}
	\theta_{\mb{q},\mb{q'}}=e^{\frac{2\pi i K_{IJ}q_Iq_J'}{n_In_J}}.
	\label{}
\end{equation}
The resolution is in the physical interpretation of \eqref{eqn:conservationlaw}: particles in general have flux lines attached to them,  so they are not really point-like objects. In fact, because of the flux lines, those particles which have nontrivial braiding statistics are actually confined, since there is generally a string tension associated with flux lines. Only when the attached flux is equal to $2\pi$(i.e. a ``Dirac string''), we have a truely deconfined quasiparticle.
The actual set of deconfined quasiparticles must have trivial mutual braiding statistics and $\pm 1$ exchange statistics.

Having in hand the physical observables, we can check the periodicity of the level $K$. From \eqref{eqn:effectiveactionfinal} we see that the effective action is completely invariant (mod $\mathbb{Z}$) under the following shift:
\begin{equation}
	\begin{aligned}
	K_{IJ}&\rightarrow K_{IJ}+n_In_J,\quad I\neq J\\
	K_{II}&\rightarrow K_{II}+2n_I^2.
	\end{aligned}
	\label{}
\end{equation}

We can also show that if $K_{II}$ is odd for any $I$, the theory contains a transparent fermionic excitation. Due to the level quantization condition $K_{II}=n_I k_{II}$, both $n_I$ and $k_{II}$ must be odd. Let us define $\vec{q}$ as $q_{J}=n_I\delta_{IJ}$. The exchange statistics of the corresponding quasiparticle is $e^{i\pi K_{II}}=-1$, i.e. fermionic. Its braiding statistics with any other quasiparticle $\mb{q}'$ reads $e^{\frac{2\pi i K_{II}q_I'}{n_I}}=1$ and obviously it has trivial braiding statistics with any loop excitations. So this is indeed a transparent fermion.

\subsection{Example: Single-component theory}
Let us consider a single component theory with $K=nk$. The exchange statistics phase of a charge-$q$ particle is $e^{\frac{i\pi k q^2 }{n}}$. Therefore, the minimal deconfined charge is $q_\text{min}=\frac{n}{\text{gcd}(k,n)}$ with exchange statistics being $\exp\left( i\pi\frac{kn}{\text{gcd}(n,k)^2} \right)=\exp\left( i\pi\frac{\text{lcm}(n,k)}{\text{gcd}(n,k)} \right)$. The action actually describes a $\mathbb{Z}_{\text{gcd}(n,k)}$ gauge theory, which is already observed in \Ref{KapustinTQFT}.

There is a nice interpretation of the minimal charge as well as the statistics. The conservation law reads 
\begin{equation}
	kj^\mu+\partial_\nu\Sigma^{\mu\nu}=0.
	\label{}
\end{equation}
We see that a charge-$q$ particle is the end of a string with flux $\frac{2\pi kq}{n}$. In order for the string to be unobservable (a Dirac string), we must have $kq/n\in \mathbb{Z}$ and the minimal $q$ is thus $q=\frac{n}{\text{gcd}(n,k)}$. The actual flux is then $2\pi\frac{k}{\text{gcd}(n,k)}$. Therefore we can consider the minimal charge as a ``dyon'' with $\frac{n}{\text{gcd}(n,k)}$ electric charges and $\frac{k}{\text{gcd}(n,k)}$ magnetic charges. The bound state has exchange statistics $(-1)^{\frac{\text{lcm}(n,k)}{\text{gcd}(n,k)} }$~\cite{dyonstat1}.

For $n$ odd, $k$ is necessarily even in a bosonic system. Therefore, $\exp\left( i\pi\frac{\text{lcm}(n,k)}{\text{gcd}(n,k)} \right)$ must be even. So all deconfined charges are bosonic.  Odd $k$ can only occur in a fermionic system. An interesting example is $k=1$ (and any odd $n$), for which the theory has fermions and no topological order (i.e. no ground state degeneracy on a torus). Thus this can be considered as a topological field theory for fermionic gapped trivial phases.

For $n$ even, the theory is necessarily bosonic and $k$ can be any integer. For example, if $k=n$ it is a $\mathbb{Z}_n$ gauge theory with a fermionic $\mathbb{Z}_n$ charge.

\section{Discussions and conclusions}
In conclusion, we establish the TQFT framework to study Abelian three-loop braiding statistics and particle braiding statistics in 3+1D. We believe such framework will be very useful for the understanding for topological phases transitions among different topological phases in 3+1D. In particular, the generalized gauge transformations defined for $B \wedge F+A\wedge A\wedge d A$ and $B \wedge F+B\wedge B$ type theory will allow use to add Maxwell terms into these theory and study dynamics beyond the topological limit. However, how to introduce matter fields to couple to 1-form and 2-form gauge fields in a gauge invariant way is still a very hard problem and will be extremely important for studying topological phases transitions via particle/loop condensations. This will be obviously an important future direction.

On the other hand, it has also been argued that $B \wedge F+A\wedge A\wedge A\wedge A$ type TQFT should describe Non-Abelian three-loop braiding statistics. In fact, similar terms have been studied in 2+1D resulting non-Abelian braiding statistics.\cite{HePRB2017} Unfortunately, the canonical quantization scheme developed in this work can not be applied to this case and much more sophisticated path integral method is needed, which is beyond the scope of this work and will be discussed else where. In the path integral quantization in spacetime, one should use a modified Wilson operator such that it is gauge invariant.

Finally, we also would like to mention that the method proposed in this work can be easily generalized into non-Abelian case as well as for interacting fermion systems after gauging fermion parity. We will discuss all these details in our future work.

\begin{acknowledgments}
We are grateful to M. Levin for very enlightening discussions. ZCG acknowledges Direct Grant no. 4053300 from The Chinese University of Hong Kong and funding from Hong Kong’s Research Grants Council (GRF no.14306918). CW was supported in part by the internal grant (No. 9610398) of City University of Hong Kong.
\end{acknowledgments}

\appendix

\section{A review of Abelian Chern-Simons particle Braiding Statistics}
In this section we briefly review the derivation of braiding and exchange statistics of quasiparticles in an Abelian Chern-Simons theory. We will carry out the calculation in two different methods, which will then be applied to topological gauge theories in $(3+1)$ dimensions.

For simplicity, we consider the following Chern-Simons theory:
\begin{equation}
	\mathcal{L}=\frac{k}{4\pi}\varepsilon^{\mu\nu\lambda}A_\mu\partial_\nu A_\lambda+j_\mu A^\mu. 
	\label{}
\end{equation}
Here $k$ is an even integer.

We first proceed using the path integral formalism and integrate out the gauge field to obtain an effective action for the current.
\begin{equation}
	\mathcal{L}_\text{eff}=\frac{2\pi}{k}\int j_\mu \frac{\epsilon^{\mu\nu\lambda}\partial_\nu}{\partial^2}j_\lambda.
	\label{}
\end{equation}

The second approach is to carry out canonical quantization for the source-free case. Take the three manifold to be $\mathcal{M}=\mathcal{M}_2\times\mathbb{R}$ where $\mathbb{R}$ corresponds to the time direction. The time components $A_{0}$ is a Lagrange multiplier and just enforce the constraint that $\varepsilon^{ij}\partial_i A_{j}=0$ when there are no external sources. We choose the temporal gauge $A_{0}=0$, and the Hilbert spaces are flat connections of $A$ modulo gauge transformations. From the canonical commutation relation we find
\begin{equation}
	[\int_\gamma A, \int_{\gamma'}A]=\frac{2\pi i}{k}\nu(\gamma,\gamma').
	\label{}
\end{equation}
Here $\nu$ is the (oriented) intersection number of the two curves $\gamma$ and $\gamma'$. Now consider the theory on a two-dimensional torus, and define the Wilson loop operators along the two non-contractible cycles:
\begin{equation}
	W_\gamma=\exp\Big(i\int_\gamma A\Big).
	\label{}
\end{equation}
Physically, one can view the Wilson loop as the following process: create a pair of anyons (i.e. particles carrying gauge charge $\pm 1$ in this case) from the ground state, and adiabatically transport the charge-$1$ particle along the cycle $\gamma$ and finally after returning to the initial position annihilate the pair.
\begin{equation}
	W_y^{-1}W_x^{-1}W_yW_x=e^{2\pi i/k}.
	\label{}
\end{equation}
This algebra of Wilson loops immediately implies the $k$-fold topological ground state degeneracy on the torus, and we can also see that the commutator corresponds to a full braid of the quasiparticles.

\section{Computation of the effective action}
\label{sec:bb}
We are ready to evaluate the effective action. We deal with $\Gamma_2$ first:
\begin{widetext}
\begin{equation}
	\begin{split}
	\Gamma_2=
	-\frac{4\pi K_{IJ}}{n_In_J}\sum_{a,b}q_{Ia}q_{Jb}\varepsilon_{\mu\nu\lambda\rho}\oint_{\gamma_a}\di y_a^\rho\oint_{\gamma_b}\di y_b^\nu\int_\mathcal{M}\di^4x\,\frac{x^\lambda-y^\lambda_a}{|x-y_a|^4}\frac{x^\mu-y^\mu_b}{|x-y_b|^4}\\
	\end{split}
	\label{}
\end{equation}
In the integral over $x$, we make a change of variable $x\rightarrow y_a+y_b-x$:
\begin{equation}
\int\di^4x\,\frac{x^\lambda-y^\lambda_a}{|x-y_a|^4}\frac{x^\mu-y^\mu_b}{|x-y_b|^4}=
\int\di^4x\,\frac{y^\lambda_b-x^\lambda}{|x-y_b|^4}\frac{y_a^\mu-x^\mu}{|x-y_a|^4}
	\label{}
\end{equation}
We immediately see that $\Gamma_2$ vanishes.

Let us turn to the more interesting term $\Gamma_1$:
	\begin{equation}
	\begin{split}
		\Gamma_1&=\int_\mathcal{M}\di^4x\,\left(\frac{2\pi K_{IJ}q_{Ia}q_{Jb}}{n_In_J}\varepsilon_{\mu\nu\lambda\rho}\Sigma_b^{\mu\nu}\frac{\partial^\lambda}{\square}j_a^\rho + \frac{2\pi}{n_I}\varepsilon_{\mu\nu\lambda\rho}{\Sigma'}_I^{\mu\nu}\frac{\partial^\lambda}{\square}j_I^\rho\right)\\
		&=\frac{K_{IJ}q_{Ia}q_{Jb}}{2\pi n_In_J}\varepsilon_{ijk}\oint_{\gamma_a}\di y_a^j\, \oint_{\gamma_b}\di y_b^i\,\int_0^\infty\di s\,\partial^k \frac{1}{|\mb{y}_a-\mb{y}_b|^2+(y_a^0-y_b^0-s)^2}+\frac{2\pi q_{Ia}\phi_{Ia} }{ n_I}I(\gamma_a, \Omega_b)\\
		&=\frac{K_{IJ}q_{Ia}q_{Jb}}{2\pi n_In_J}\varepsilon_{ijk}\oint_{\gamma_a}\di y_a^j\, \oint_{\gamma_b}\di y_b^i\, \frac{y_a^k-y_b^k}{|\mb{y}_a-\mb{y}_b|^3}\left( \frac{\pi}{2}+\arctan\frac{y_a^0-y_b^0}{|\mb{y}_a-\mb{y}_b|} + \frac{ (y_a^0-y_b^0)|\mb{y}_a-\mb{y}_b|}{|y_a-y_b|^2} \right)+\frac{2\pi q_{Ia}\phi_{Ia} }{ n_I}I(\gamma_a, \Omega_b).
	\end{split}
	\label{}
\end{equation}
\end{widetext}
Here we have chosen $u=(1,0,0,0)$. 
$I(\gamma,\Omega)$ is the linking number between the surface $\Omega$ and the curve $\gamma$, which represents the familar Aharonov-Boham phase between charged particles and flux loops in three dimensions.

The remaining integral naturally separates into a part that is symmetric under the interchange of $a$ and $b$(i.e. the $\frac{\pi}{2}$ term), and one that is anti-symmetric. 
First we consider terms with $a\neq b$. In this case, the anti-symmetric part cancels out, and the symmetric part just evaluates to $\frac{\pi K_{IJ}q_{Ia}q_{Jb}}{n_In_J}I(\gamma_a,\gamma_b)$ where $I(\gamma_a,\gamma_b)$ is the linking number of the 3D projection of $\gamma_a$ and $\gamma_b$:
\begin{equation}
	I(\gamma_a,\gamma_b)=\frac{1}{4\pi}\varepsilon_{ijk}\oint_{\gamma_a}\di y_a^i\, \oint_{\gamma_b}\di y_b^j\, \frac{y_a^k-y_b^k}{|\mb{y}_a-\mb{y}_b|^3}.
	\label{}
\end{equation}
For $a=b$, the integral as given is not well-defined and needs to be regularized. A standard regularization for such integral can by done by the procedure of framing, in which one displaces the curve appearing in the first argument by an infinitesimal $3$-vector field normal to the curve. Since the $0$-components stay the same, the ``anti-symmetric'' part vanishes. The result of the integral is just $\frac{\pi K_{IJ}q_{Ia}q_{Jb}}{n_In_J}I(\gamma_a)$ where $I(\gamma_a)$ is now the self-linking number of the 3D projection of $\gamma_a$.

\section{A microscopic derivation for the gauge transformation of $A\wedge A \wedge dA$ type theory}
Let us motivate the field theory and the generalized gauge transformations using a microscopic model. Consider a bosonic superconductor. There are $N$ $\mathrm{U}(1)$ phases $\theta_I$, carrying $n_I$ charges respectively, coupled to (compact) gauge fields $A_I$. The Lagrangian is the standard Abelian Higgs model plus a nonlinear interaction between current and gauge field strength:
\begin{equation}
	\begin{split}
		\mathcal{L}=&\frac{1}{2\rho}(\partial_\mu \theta_I-n_IA_{I\mu})^2 + \frac{1}{g^2}F_{I\mu\nu}^2 \\
		&+ i\Lambda_{IJK}\varepsilon^{\mu\nu\lambda\rho}(\partial_\mu \theta_I-n_IA_{I\mu})(\partial_\nu\theta_J-n_JA_{J\nu})F_{K\lambda\rho}
	\end{split}
	\label{}
\end{equation}
Without loss of generality, we can set $\Lambda_{IJK}=-\Lambda_{JIK}$, where $\Lambda_{IJK}=\frac{M_{IJK}}{8\pi^2 n_In_J}$.
We now apply the standard duality transformation to the model. We separate the phase field to the smooth and singular (i.e. vortex lines) part: $\theta_I=\theta_I^s+\theta_I^v$, and introduce Hubbard-Stratonovich fields:
\begin{widetext}
\begin{equation}
	\begin{split}
	\mathcal{L}=&\frac{1}{2\rho}\xi_I^2+i\xi_I^\mu(\partial_\mu\theta^s_{I}-n_IA_{I\mu})
	- in_J\Lambda_{IJK}\varepsilon^{\mu\nu\lambda\rho}\partial_\mu \theta_I^sA_{J\nu}\partial_\lambda A_{K\rho}
	+ in_I\Lambda_{IJK}\varepsilon^{\mu\nu\lambda\rho}\partial_\mu \theta_J^sA_{I\nu}\partial_\lambda A_{K\rho}\\
&+in_In_J\Lambda_{IJK} \varepsilon^{\mu\nu\lambda\rho} A_{I\mu}A_{J\nu}\partial_\lambda A_{K\rho}
+in_In_J\Lambda_{IJK} \varepsilon^{\mu\nu\lambda\rho} \partial_\mu\theta_I^s\partial_\nu\theta_J^s\partial_\lambda A_{K\rho}
	\end{split}
	\label{}
\end{equation}
\end{widetext}
The last term can be written as a total derivative and we will drop it in the following.
Integrating out the smooth part $\theta_I^s$, we find the following constraint
\begin{equation}
	\begin{gathered}
		\partial_\mu\Big[\xi_I^\mu- \tilde{\Lambda}_{IJK}\varepsilon^{\mu\nu\lambda\rho} A_{J\nu}\partial_\lambda A_{K\rho}\Big]=0
	\end{gathered}
	\label{}
\end{equation}
Here we have defined $\tilde{\Lambda}_{IJK}=n_J(\Lambda_{JIK} - \Lambda_{IJK})$ for brevity. We can introduce a (formally non-compact) $2$-form gauge field $B_{I\mu\nu}$ to resolve the constraint:
\begin{equation}
	\begin{gathered}
		\xi_I^\mu=\tilde{\Lambda}_{IJK}\varepsilon^{\mu\nu\lambda\rho}A_{J\nu}\partial_\lambda A_{K\rho}-\frac{\varepsilon^{\mu\nu\lambda\rho}}{4\pi}\partial_\nu B_{I\lambda\rho}
	\end{gathered}
	\label{}
\end{equation}
From here we can derive the gauge transformation of $B$. Since $\xi_I$ should be gauge-invariant, $B$ should transform in such a way to cancel the gauge transformation of the first term.
\begin{equation}
	A_{I\mu}\rightarrow A_{I\mu}+\partial_\mu f_I, B_I\rightarrow B_I + 4\pi \tilde{\Lambda}_{IJK}f_J\partial_\lambda A_{K\rho}
	\label{}
\end{equation}
The dual action then reads
\begin{equation}
	\mathcal{L}=\frac{n_I}{4\pi}\varepsilon^{\mu\nu\lambda\rho}A_{I\mu}\partial_\nu B_{I\lambda\rho} -n_In_J\Lambda_{IJK}A_{I\mu}A_{J\nu}\partial_\lambda A_{K\rho}.
	\label{}
\end{equation}

\section{Equivalent actions}
\label{App:equi}
Up to a total derivative, the second term in the action \eq{eqn:dw} can be  written as
\begin{align}
\frac{M_{IJK}}{8\pi^2} A_IA_J\di A_K &= -\frac{M_{IJK}}{8\pi^2} \di(A_IA_J) A_K\\\nonumber
&= -\frac{2M_{KIJ}}{8\pi^2} A_IA_J\di A_K\\\nonumber
&= \frac{M_{KJI}-M_{KIJ}}{8\pi^2} A_IA_J\di A_K,
\end{align}
where we have used $M_{IJK}=-M_{JIK}$ and $A_IA_J=-A_JA_I$. Therefore, two actions of the form \eq{eqn:dw} with $M_{IJK}$ and $M_{IJK}'$, where
\begin{align}
M_{IJK}'=M_{KJI}-M_{KIJ},
\end{align}
should be equivalent. In particular, the three-loop braiding statistics, as a gauge invariant physical quantity, should be the same for the two theories.

Using the three-loop braiding statistics result \eq{eq:3loop}, one can check
\begin{align}\nonumber
e^{i\theta_{IJ,K}'}
&= \exp\left[ {-\frac{4\pi i}{n_I n_J n_K} (M_{IKJ}'+M_{JKI}') }\right]\\\nonumber
&= \exp\left[ {-\frac{4\pi i}{n_I n_J n_K} (M_{JKI}-M_{JIK}+M_{IKJ}-M_{IJK}) }\right]\\\nonumber
&= \exp\left[ {-\frac{4\pi i}{n_I n_J n_K} (M_{JKI}+M_{IKJ}) }\right]\\
&= e^{i\theta_{IJ,K}},
\end{align}
where we used again $M_{IJK}=-M_{JIK}$. Therefore, the theories with $M_{IJK}$ and $M_{IJK}'$ indeed have the same three-loop braiding statistics.

\bibliography{spt}

\begin{thebibliography}{39}
\expandafter\ifx\csname natexlab\endcsname\relax\def\natexlab#1{#1}\fi
\expandafter\ifx\csname bibnamefont\endcsname\relax
  \def\bibnamefont#1{#1}\fi
\expandafter\ifx\csname bibfnamefont\endcsname\relax
  \def\bibfnamefont#1{#1}\fi
\expandafter\ifx\csname citenamefont\endcsname\relax
  \def\citenamefont#1{#1}\fi
\expandafter\ifx\csname url\endcsname\relax
  \def\url#1{\texttt{#1}}\fi
\expandafter\ifx\csname urlprefix\endcsname\relax\def\urlprefix{URL }\fi
\providecommand{\bibinfo}[2]{#2}
\providecommand{\eprint}[2][]{\url{#2}}

\bibitem[{\citenamefont{Wen}(2004)}]{wen-book}
\bibinfo{author}{\bibfnamefont{X.-G.} \bibnamefont{Wen}},
  \emph{\bibinfo{title}{Quantum Field Theory of Many-Body Systems}}
  (\bibinfo{publisher}{Oxford University Press}, \bibinfo{address}{Oxford},
  \bibinfo{year}{2004}).

\bibitem[{\citenamefont{Nayak et~al.}(2008)\citenamefont{Nayak, Simon, Stern,
  Freedman, and Sarma}}]{Nayak08}
\bibinfo{author}{\bibfnamefont{C.}~\bibnamefont{Nayak}},
  \bibinfo{author}{\bibfnamefont{S.~H.} \bibnamefont{Simon}},
  \bibinfo{author}{\bibfnamefont{A.}~\bibnamefont{Stern}},
  \bibinfo{author}{\bibfnamefont{M.}~\bibnamefont{Freedman}}, \bibnamefont{and}
  \bibinfo{author}{\bibfnamefont{S.~D.} \bibnamefont{Sarma}},
  \bibinfo{journal}{Rev. Mod. Phys.} \textbf{\bibinfo{volume}{80}},
  \bibinfo{pages}{1083} (\bibinfo{year}{2008}), \eprint{arXiv:0707.1889}.

\bibitem[{\citenamefont{Dijkgraaf and Witten}(1990)}]{Dijkgraaf90}
\bibinfo{author}{\bibfnamefont{R.}~\bibnamefont{Dijkgraaf}} \bibnamefont{and}
  \bibinfo{author}{\bibfnamefont{E.}~\bibnamefont{Witten}},
  \bibinfo{journal}{Commun. Math. Phys.} \textbf{\bibinfo{volume}{129}},
  \bibinfo{pages}{393} (\bibinfo{year}{1990}).

\bibitem[{\citenamefont{Wang and Levin}(2014)}]{Levin_PRL2014}
\bibinfo{author}{\bibfnamefont{C.}~\bibnamefont{Wang}} \bibnamefont{and}
  \bibinfo{author}{\bibfnamefont{M.}~\bibnamefont{Levin}},
  \bibinfo{journal}{Phys. Rev. Lett.} \textbf{\bibinfo{volume}{113}},
  \bibinfo{pages}{080403} (\bibinfo{year}{2014}).

\bibitem[{\citenamefont{{Wang} and {Levin}}(2015)}]{Wang_arxiv2014}
\bibinfo{author}{\bibfnamefont{C.}~\bibnamefont{{Wang}}} \bibnamefont{and}
  \bibinfo{author}{\bibfnamefont{M.}~\bibnamefont{{Levin}}},
  \bibinfo{journal}{\prb} \textbf{\bibinfo{volume}{91}}, \bibinfo{eid}{165119}
  (\bibinfo{year}{2015}), \eprint{1412.1781}.

\bibitem[{\citenamefont{Jiang et~al.}(2014)\citenamefont{Jiang, Mesaros, and
  Ran}}]{Ran_PRX2014}
\bibinfo{author}{\bibfnamefont{S.}~\bibnamefont{Jiang}},
  \bibinfo{author}{\bibfnamefont{A.}~\bibnamefont{Mesaros}}, \bibnamefont{and}
  \bibinfo{author}{\bibfnamefont{Y.}~\bibnamefont{Ran}},
  \bibinfo{journal}{Phys. Rev. X} \textbf{\bibinfo{volume}{4}},
  \bibinfo{pages}{031048} (\bibinfo{year}{2014}).

\bibitem[{\citenamefont{Wang and Wen}(2015)}]{Wang_DW}
\bibinfo{author}{\bibfnamefont{J.~C.} \bibnamefont{Wang}} \bibnamefont{and}
  \bibinfo{author}{\bibfnamefont{X.-G.} \bibnamefont{Wen}},
  \bibinfo{journal}{Phys. Rev. B} \textbf{\bibinfo{volume}{91}},
  \bibinfo{pages}{035134} (\bibinfo{year}{2015}).

\bibitem[{\citenamefont{{Wan} et~al.}(2015)\citenamefont{{Wan}, {Wang}, and
  {He}}}]{Wan_DW}
\bibinfo{author}{\bibfnamefont{Y.}~\bibnamefont{{Wan}}},
  \bibinfo{author}{\bibfnamefont{J.~C.} \bibnamefont{{Wang}}},
  \bibnamefont{and} \bibinfo{author}{\bibfnamefont{H.}~\bibnamefont{{He}}},
  \bibinfo{journal}{\prb} \textbf{\bibinfo{volume}{92}}, \bibinfo{eid}{045101}
  (\bibinfo{year}{2015}), \eprint{1409.3216}.

\bibitem[{\citenamefont{Moradi and Wen}(2015)}]{Moradi_arxiv2014}
\bibinfo{author}{\bibfnamefont{H.}~\bibnamefont{Moradi}} \bibnamefont{and}
  \bibinfo{author}{\bibfnamefont{X.-G.} \bibnamefont{Wen}},
  \bibinfo{journal}{Phys. Rev. B} \textbf{\bibinfo{volume}{91}},
  \bibinfo{pages}{075114} (\bibinfo{year}{2015}).

\bibitem[{\citenamefont{Jian and Qi}(2014)}]{Jian_unpub}
\bibinfo{author}{\bibfnamefont{C.-M.} \bibnamefont{Jian}} \bibnamefont{and}
  \bibinfo{author}{\bibfnamefont{X.-L.} \bibnamefont{Qi}},
  \bibinfo{journal}{Phys. Rev. X} \textbf{\bibinfo{volume}{4}},
  \bibinfo{pages}{041043} (\bibinfo{year}{2014}), \eprint{arXiv:1405.6688}.

\bibitem[{\citenamefont{Bi et~al.}(2014)\citenamefont{Bi, You, and
  Xu}}]{Bi_PRB2014}
\bibinfo{author}{\bibfnamefont{Z.}~\bibnamefont{Bi}},
  \bibinfo{author}{\bibfnamefont{Y.-Z.} \bibnamefont{You}}, \bibnamefont{and}
  \bibinfo{author}{\bibfnamefont{C.}~\bibnamefont{Xu}}, \bibinfo{journal}{Phys.
  Rev. B} \textbf{\bibinfo{volume}{90}}, \bibinfo{pages}{081110}
  (\bibinfo{year}{2014}).

\bibitem[{\citenamefont{Walker and Wang}(2012)}]{WW}
\bibinfo{author}{\bibfnamefont{K.}~\bibnamefont{Walker}} \bibnamefont{and}
  \bibinfo{author}{\bibfnamefont{Z.}~\bibnamefont{Wang}},
  \bibinfo{journal}{Frontier of Physics} \textbf{\bibinfo{volume}{7}},
  \bibinfo{pages}{150} (\bibinfo{year}{2012}).

\bibitem[{\citenamefont{von Keyserlingk et~al.}(2013)\citenamefont{von
  Keyserlingk, Burnell, and Simon}}]{CurtPRB2013}
\bibinfo{author}{\bibfnamefont{C.~W.} \bibnamefont{von Keyserlingk}},
  \bibinfo{author}{\bibfnamefont{F.~J.} \bibnamefont{Burnell}},
  \bibnamefont{and} \bibinfo{author}{\bibfnamefont{S.~H.} \bibnamefont{Simon}},
  \bibinfo{journal}{Phys. Rev. B} \textbf{\bibinfo{volume}{87}},
  \bibinfo{pages}{045107} (\bibinfo{year}{2013}).

\bibitem[{\citenamefont{{von Keyserlingk} and {Burnell}}(2015)}]{Curt_unpub}
\bibinfo{author}{\bibfnamefont{C.~W.} \bibnamefont{{von Keyserlingk}}}
  \bibnamefont{and} \bibinfo{author}{\bibfnamefont{F.~J.}
  \bibnamefont{{Burnell}}}, \bibinfo{journal}{\prb}
  \textbf{\bibinfo{volume}{91}}, \bibinfo{eid}{045134} (\bibinfo{year}{2015}),
  \eprint{1405.2988}.

\bibitem[{\citenamefont{{Putrov} et~al.}(2017)\citenamefont{{Putrov}, {Wang},
  and {Yau}}}]{PutrovAP2017}
\bibinfo{author}{\bibfnamefont{P.}~\bibnamefont{{Putrov}}},
  \bibinfo{author}{\bibfnamefont{J.}~\bibnamefont{{Wang}}}, \bibnamefont{and}
  \bibinfo{author}{\bibfnamefont{S.-T.} \bibnamefont{{Yau}}},
  \bibinfo{journal}{Annals of Physics} \textbf{\bibinfo{volume}{384}},
  \bibinfo{pages}{254} (\bibinfo{year}{2017}), \eprint{1612.09298}.

\bibitem[{\citenamefont{Chen et~al.}(2013)\citenamefont{Chen, Gu, Liu, and
  Wen}}]{Chen_arxiv2011}
\bibinfo{author}{\bibfnamefont{X.}~\bibnamefont{Chen}},
  \bibinfo{author}{\bibfnamefont{Z.-C.} \bibnamefont{Gu}},
  \bibinfo{author}{\bibfnamefont{Z.-X.} \bibnamefont{Liu}}, \bibnamefont{and}
  \bibinfo{author}{\bibfnamefont{X.-G.} \bibnamefont{Wen}},
  \bibinfo{journal}{Phys. Rev. B} \textbf{\bibinfo{volume}{87}},
  \bibinfo{pages}{155114} (\bibinfo{year}{2013}).

\bibitem[{\citenamefont{Levin and Gu}(2012)}]{LevinGu_arxiv2012}
\bibinfo{author}{\bibfnamefont{M.}~\bibnamefont{Levin}} \bibnamefont{and}
  \bibinfo{author}{\bibfnamefont{Z.-C.} \bibnamefont{Gu}},
  \bibinfo{journal}{Phys. Rev. B} \textbf{\bibinfo{volume}{86}},
  \bibinfo{pages}{115109} (\bibinfo{year}{2012}).

\bibitem[{\citenamefont{Cheng and Gu}(2014)}]{ChengPRL2014}
\bibinfo{author}{\bibfnamefont{M.}~\bibnamefont{Cheng}} \bibnamefont{and}
  \bibinfo{author}{\bibfnamefont{Z.-C.} \bibnamefont{Gu}},
  \bibinfo{journal}{Phys. Rev. Lett.} \textbf{\bibinfo{volume}{112}},
  \bibinfo{pages}{141602} (\bibinfo{year}{2014}).

\bibitem[{\citenamefont{Cheng et~al.}(2018)\citenamefont{Cheng,
  Tantivasadakarn, and Wang}}]{ChengPRX2018}
\bibinfo{author}{\bibfnamefont{M.}~\bibnamefont{Cheng}},
  \bibinfo{author}{\bibfnamefont{N.}~\bibnamefont{Tantivasadakarn}},
  \bibnamefont{and} \bibinfo{author}{\bibfnamefont{C.}~\bibnamefont{Wang}},
  \bibinfo{journal}{Phys. Rev. X} \textbf{\bibinfo{volume}{8}},
  \bibinfo{pages}{011054} (\bibinfo{year}{2018}).

\bibitem[{\citenamefont{Wang et~al.}(2015)\citenamefont{Wang, Gu, and
  Wen}}]{Wang_unpub}
\bibinfo{author}{\bibfnamefont{J.~C.} \bibnamefont{Wang}},
  \bibinfo{author}{\bibfnamefont{Z.-C.} \bibnamefont{Gu}}, \bibnamefont{and}
  \bibinfo{author}{\bibfnamefont{X.-G.} \bibnamefont{Wen}},
  \bibinfo{journal}{Phys. Rev. Lett.} \textbf{\bibinfo{volume}{114}},
  \bibinfo{pages}{031601} (\bibinfo{year}{2015}).

\bibitem[{\citenamefont{Lan et~al.}(2017)\citenamefont{Lan, Kong, and
  Wen}}]{tian2017}
\bibinfo{author}{\bibfnamefont{T.}~\bibnamefont{Lan}},
  \bibinfo{author}{\bibfnamefont{L.}~\bibnamefont{Kong}}, \bibnamefont{and}
  \bibinfo{author}{\bibfnamefont{X.-G.} \bibnamefont{Wen}},
  \bibinfo{journal}{Phys. Rev. X} \textbf{\bibinfo{volume}{8}},
  \bibinfo{pages}{021074} (\bibinfo{year}{2017}).

\bibitem[{\citenamefont{Deligne}(2002)}]{Deligne}
\bibinfo{author}{\bibfnamefont{P.}~\bibnamefont{Deligne}},
  \bibinfo{journal}{Moscow Math. Journal} \textbf{\bibinfo{volume}{2}},
  \bibinfo{pages}{227} (\bibinfo{year}{2002}).

\bibitem[{\citenamefont{Haah}(2011)}]{Haah}
\bibinfo{author}{\bibfnamefont{J.}~\bibnamefont{Haah}}, \bibinfo{journal}{Phys.
  Rev. A} \textbf{\bibinfo{volume}{83}}, \bibinfo{pages}{042330}
  (\bibinfo{year}{2011}),
  \urlprefix\url{https://link.aps.org/doi/10.1103/PhysRevA.83.042330}.

\bibitem[{\citenamefont{Vijay et~al.}(2015)\citenamefont{Vijay, Haah, and
  Fu}}]{VijayPRB2015}
\bibinfo{author}{\bibfnamefont{S.}~\bibnamefont{Vijay}},
  \bibinfo{author}{\bibfnamefont{J.}~\bibnamefont{Haah}}, \bibnamefont{and}
  \bibinfo{author}{\bibfnamefont{L.}~\bibnamefont{Fu}}, \bibinfo{journal}{Phys.
  Rev. B} \textbf{\bibinfo{volume}{92}}, \bibinfo{pages}{235136}
  (\bibinfo{year}{2015}).

\bibitem[{\citenamefont{Vijay et~al.}(2016)\citenamefont{Vijay, Haah, and
  Fu}}]{VijayPRB2016}
\bibinfo{author}{\bibfnamefont{S.}~\bibnamefont{Vijay}},
  \bibinfo{author}{\bibfnamefont{J.}~\bibnamefont{Haah}}, \bibnamefont{and}
  \bibinfo{author}{\bibfnamefont{L.}~\bibnamefont{Fu}}, \bibinfo{journal}{Phys.
  Rev. B} \textbf{\bibinfo{volume}{94}}, \bibinfo{pages}{235157}
  (\bibinfo{year}{2016}).

\bibitem[{\citenamefont{{Kapustin} and {Thorngren}}(2013)}]{KapustinArxiv2013}
\bibinfo{author}{\bibfnamefont{A.}~\bibnamefont{{Kapustin}}} \bibnamefont{and}
  \bibinfo{author}{\bibfnamefont{R.}~\bibnamefont{{Thorngren}}},
  \bibinfo{journal}{ArXiv e-prints}  (\bibinfo{year}{2013}),
  \eprint{1308.2926}.

\bibitem[{\citenamefont{{Kapustin} and {Thorngren}}(2014)}]{kapustin2014}
\bibinfo{author}{\bibfnamefont{A.}~\bibnamefont{{Kapustin}}} \bibnamefont{and}
  \bibinfo{author}{\bibfnamefont{R.}~\bibnamefont{{Thorngren}}},
  \bibinfo{journal}{ArXiv e-prints}  (\bibinfo{year}{2014}),
  \eprint{1404.3230}.

\bibitem[{\citenamefont{{Ye} and {Gu}}(2016)}]{Ye_unpub2015}
\bibinfo{author}{\bibfnamefont{P.}~\bibnamefont{{Ye}}} \bibnamefont{and}
  \bibinfo{author}{\bibfnamefont{Z.-C.} \bibnamefont{{Gu}}},
  \bibinfo{journal}{\prb} \textbf{\bibinfo{volume}{93}}, \bibinfo{eid}{205157}
  (\bibinfo{year}{2016}), \eprint{1508.05689}.

\bibitem[{\citenamefont{{Chen} et~al.}(2016)\citenamefont{{Chen}, {Tiwari}, and
  {Ryu}}}]{Ryu2015}
\bibinfo{author}{\bibfnamefont{X.}~\bibnamefont{{Chen}}},
  \bibinfo{author}{\bibfnamefont{A.}~\bibnamefont{{Tiwari}}}, \bibnamefont{and}
  \bibinfo{author}{\bibfnamefont{S.}~\bibnamefont{{Ryu}}},
  \bibinfo{journal}{\prb} \textbf{\bibinfo{volume}{94}}, \bibinfo{eid}{045113}
  (\bibinfo{year}{2016}), \eprint{1509.04266}.

\bibitem[{\citenamefont{Tiwari et~al.}(2017)\citenamefont{Tiwari, Chen, and
  Ryu}}]{TiwariPRB2017}
\bibinfo{author}{\bibfnamefont{A.}~\bibnamefont{Tiwari}},
  \bibinfo{author}{\bibfnamefont{X.}~\bibnamefont{Chen}}, \bibnamefont{and}
  \bibinfo{author}{\bibfnamefont{S.}~\bibnamefont{Ryu}},
  \bibinfo{journal}{Phys. Rev. B} \textbf{\bibinfo{volume}{95}},
  \bibinfo{pages}{245124} (\bibinfo{year}{2017}).

\bibitem[{\citenamefont{{Gu} et~al.}(2016)\citenamefont{{Gu}, {Wang}, and
  {Wen}}}]{Gu_unpub}
\bibinfo{author}{\bibfnamefont{Z.-C.} \bibnamefont{{Gu}}},
  \bibinfo{author}{\bibfnamefont{J.~C.} \bibnamefont{{Wang}}},
  \bibnamefont{and} \bibinfo{author}{\bibfnamefont{X.-G.} \bibnamefont{{Wen}}},
  \bibinfo{journal}{\prb} \textbf{\bibinfo{volume}{93}}, \bibinfo{eid}{115136}
  (\bibinfo{year}{2016}), \eprint{1503.01768}.

\bibitem[{\citenamefont{{Chan} et~al.}(2018)\citenamefont{{Chan}, {Ye}, and
  {Ryu}}}]{chan17}
\bibinfo{author}{\bibfnamefont{A.~P.~O.} \bibnamefont{{Chan}}},
  \bibinfo{author}{\bibfnamefont{P.}~\bibnamefont{{Ye}}}, \bibnamefont{and}
  \bibinfo{author}{\bibfnamefont{S.}~\bibnamefont{{Ryu}}},
  \bibinfo{journal}{Physical Review Letters} \textbf{\bibinfo{volume}{121}},
  \bibinfo{eid}{061601} (\bibinfo{year}{2018}), \eprint{1703.01926}.

\bibitem[{\citenamefont{{Yoshida}}(2017)}]{Yoshida}
\bibinfo{author}{\bibfnamefont{B.}~\bibnamefont{{Yoshida}}},
  \bibinfo{journal}{Annals of Physics} \textbf{\bibinfo{volume}{377}},
  \bibinfo{pages}{387} (\bibinfo{year}{2017}), \eprint{1509.03626}.

\bibitem[{\citenamefont{Horowitz}(1989)}]{HorowitzCMP}
\bibinfo{author}{\bibfnamefont{G.~T.} \bibnamefont{Horowitz}},
  \bibinfo{journal}{Commun. Math. Phys.} \textbf{\bibinfo{volume}{125}},
  \bibinfo{pages}{417} (\bibinfo{year}{1989}).

\bibitem[{\citenamefont{Ye and Gu}(2015)}]{Ye_unpub}
\bibinfo{author}{\bibfnamefont{P.}~\bibnamefont{Ye}} \bibnamefont{and}
  \bibinfo{author}{\bibfnamefont{Z.-C.} \bibnamefont{Gu}},
  \bibinfo{journal}{Phys. Rev. X} \textbf{\bibinfo{volume}{5}},
  \bibinfo{pages}{021029} (\bibinfo{year}{2015}).

\bibitem[{\citenamefont{{Lechner} and {Marchetti}}(2000)}]{LechnerJHEP2000}
\bibinfo{author}{\bibfnamefont{K.}~\bibnamefont{{Lechner}}} \bibnamefont{and}
  \bibinfo{author}{\bibfnamefont{P.~A.} \bibnamefont{{Marchetti}}},
  \bibinfo{journal}{Journal of High Energy Physics}
  \textbf{\bibinfo{volume}{12}}, \bibinfo{pages}{028} (\bibinfo{year}{2000}),
  \eprint{hep-th/0010291}.

\bibitem[{\citenamefont{Kapustin and Seiberg}()}]{KapustinTQFT}
\bibinfo{author}{\bibfnamefont{A.}~\bibnamefont{Kapustin}} \bibnamefont{and}
  \bibinfo{author}{\bibfnamefont{N.}~\bibnamefont{Seiberg}},
  \eprint{arXiv:1401.0740}.

\bibitem[{\citenamefont{Goldhaber}(1976)}]{dyonstat1}
\bibinfo{author}{\bibfnamefont{A.~S.} \bibnamefont{Goldhaber}},
  \bibinfo{journal}{Phys. Rev. Lett.} \textbf{\bibinfo{volume}{36}},
  \bibinfo{pages}{1122} (\bibinfo{year}{1976}).

\bibitem[{\citenamefont{He et~al.}(2017)\citenamefont{He, Zheng, and von
  Keyserlingk}}]{HePRB2017}
\bibinfo{author}{\bibfnamefont{H.}~\bibnamefont{He}},
  \bibinfo{author}{\bibfnamefont{Y.}~\bibnamefont{Zheng}}, \bibnamefont{and}
  \bibinfo{author}{\bibfnamefont{C.}~\bibnamefont{von Keyserlingk}},
  \bibinfo{journal}{Phys. Rev. B} \textbf{\bibinfo{volume}{95}},
  \bibinfo{pages}{035131} (\bibinfo{year}{2017}).

\end{thebibliography}

\end{document}